\documentclass[journal=jpclcd,manuscript=letter,layout=twocolumn]{achemso}

\usepackage{amsfonts,bm}
\usepackage{dcolumn}
\usepackage{graphicx}
\usepackage{import}
\usepackage{nicematrix}
\usepackage{soul}
\usepackage{tikz}
\usepackage[normalem]{ulem}
\usepackage{xcolor}
\usepackage[colorlinks=true,linkcolor=blue,citecolor=blue]{hyperref}
\usepackage[capitalise,nameinlink]{cleveref}

\def\meuwly{CCSD(T)-5$~$}
\def\he{MRCI-5$~$}
\def\ma{MRCI-4$~$}
\def\wavenum{\,\mathrm{cm}^{-1}}

\newcommand{\change}[1]{{\color{black} #1}}

\title{Feshbach Resonances in Cold Collisions: \\ Benchmarking
  State of the Art \textit{ab initio} Potential Energy Surfaces}

\author{Karl P. Horn}
\affiliation{Freie Universit{\"a}t Berlin, 
  Dahlem Center for Complex Quantum Systems and Fachbereich Physik,
  Arnimallee 14, 14195 Berlin, Germany}
\author{Meenu Upadhyay}
\affiliation{Department of Chemistry, University of Basel, Basel, Switzerland.}
\author{Baruch Margulis}
\affiliation{Department of Chemical and Biological Physics, Weizmann Institute of Science, 7610001 Rehovot, Israel.}
\altaffiliation{{Current address: National Institute of Standards and Technologies, 80305 Boulder, CO, USA.}}
\author{Daniel M. Reich}
\affiliation{Freie Universit{\"a}t Berlin, 
  Dahlem Center for Complex Quantum Systems and Fachbereich Physik,
  Arnimallee 14, 14195 Berlin, Germany}
\author{Edvardas Narevicius}
\affiliation{Department of Chemical and Biological Physics, Weizmann Institute of Science, 7610001 Rehovot, Israel}
\affiliation{Department of Physics, Technische Universit\"at Dortmund, Dortmund, Germany}
\author{Markus Meuwly}
\affiliation{Department of Chemistry, University of Basel, Basel, Switzerland}
\author{Christiane P. Koch}
\email{christiane.koch@fu-berlin.de}
\affiliation{  Freie Universit{\"a}t Berlin,
  Dahlem Center for Complex Quantum Systems and Fachbereich Physik,
  Arnimallee 14, 14195 Berlin, Germany}

\begin{tocentry}
  \includegraphics[width=5cm]{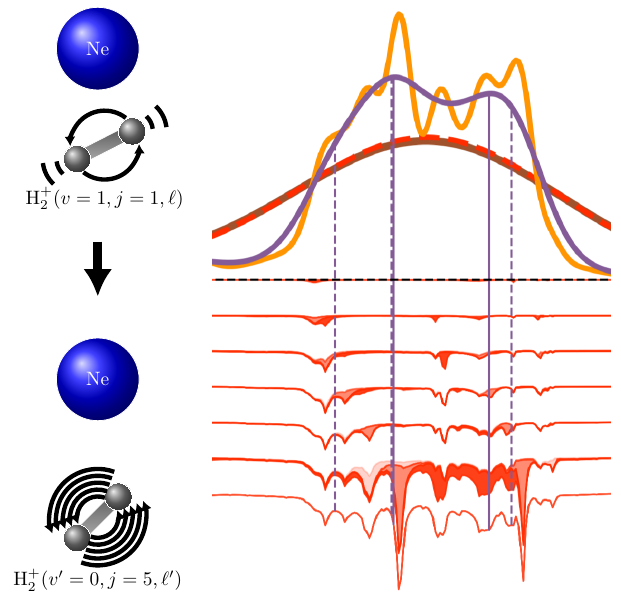}
  How far are cold collision experiments from benchmarking the best potential energy surfaces?
\end{tocentry}

\mciteErrorOnUnknownfalse
\begin{document}
\begin{abstract}
High-quality potential energy surfaces (PES) are a prerequisite for
quantitative atomistic simulations, with both quantum and classical
dynamics approaches. The ultimate test for the validity of a PES are
comparisons with judiciously chosen experimental observables.
Here we ask whether cold collision measurements are sufficiently informative to validate and distinguish between high-level, state-of-the art PESs for the strongly interacting Ne--H$_2^+$ system.  We show that measurement of the final state distributions for a process that involves only  several metastable intermediate 
states  is sufficient to identify  the PES that captures the long-range interactions properly. 
Furthermore, we show that a modest increase in the experimental energy resolution will allow for resolving individual Feshbach resonances and enable a quantitative probe of the interactions at short and intermediate range.
\end{abstract}

\maketitle

Collision experiments measure the probability of the collision
partners to change their internal state(s) or undergo a chemical
reaction\cite{JohnsonBook}. Since collisions ultimately probe
inter-particle interactions, comparing measured and calculated
collision cross sections allows us to quantify, in principle, how well
our theoretical understanding matches physical reality in the
experiments\cite{FriedrichBook,KarmanReview}. For molecular
collisions, however, such a quantitative assessment of theoretical
models has long been hampered by the stringent requirement that
experiments need to resolve both initial and final
states\cite{YangARPC2007,JankunasARPC}. Investigating cold and
ultracold collisions provide a means to prepare well-defined initial
states with quantum purity\cite{DulieuOsterwalderBook}. Final-state
resolution has recently been added to these experiments by collecting
the reaction products with velocity map
imaging\cite{liu2021precision,margulis2023tomography,TangSci2023,PlompJPCL2024}. But
even with state-to-state resolution it is often challenging to
identify suitable experimental observables that provide information on
local or global properties of the underlying potential energy surface (PES).\cite{MoritaPRL2019}
This is true in particular for collision systems where the interaction
is strong enough to couple different internal and external degrees of
freedom\cite{liu2021precision,ZhouNatChem2022,deJonghNatChem2022,ManPRL2022}.

For molecular collisions, the interaction between the partners is
fully described by the PES. How much information about a PES
can be inferred from measurements depends on 
characteristics such as the range and degree of anisotropy of the
intermolecular interactions. For example, an almost fully statistical
distribution over the energetically allowed final states was observed
in reactive collisions with very strong
anisotropy~\cite{liu2021precision}, even though the collisions proceed in
the fully quantum regime. In contrast, clear quantum fingerprints in
the cross sections have been observed in reactive collisions between
rare gas atoms and dihydrogen molecules, even though the interaction is
similarly anisotropic\cite{margulis2023tomography}. This difference points to the importance of resolving not only initial and final, but also intermediate quantum states of a collision.

Quantum scattering calculations for atom-diatom complexes as considered here can be fully converged. Thus, the computed
observables only depend on the quality of the underlying PES
\change{which is determined by the electronic structure method  and numerical technique used to calculate and represent it.}
Changes in the shape of the PES will directly translate into modifications of the
observables which can be exploited, for example, by ``PES
morphing'' through suitable coordinate
transformations.\cite{MeuwlyJCP1999,horn2023improving,bowman:1991}
\change{At the same time, a given physical observable is usually sensitive only to specific regions and properties of a PES. For example, Feshbach resonances are particularly sensitive to and informative about the long-range part of
  a PES but do not  directly provide information about the well depth.
  Even with an exhaustive set of measured Feshbach resonances not all properties of a full-dimensional PES can be probed and adjusted.\cite{horn2023improving}}

In the present contribution we invert this perspective by asking:
Can we distinguish the quality of different PESs and learn about
their merits and deficiencies? By how much does the
experimental resolution need to be improved to discriminate between
different PESs? The proxy for addressing these questions is the
Ne--H$_2^+$ complex for which an exhaustive list of measured Feshbach
resonances is available.\cite{margulis2023tomography} 
  These metastable intermediate states
probe wide parts of the PES;\cite{horn2023improving} they are ubiquitous in few-body reaction dynamics\cite{otto2014,ma2015,zhang2020,liu2023,park2023} and were shown to lead to unique quantum  fingerprints in the final state distribution.\cite{margulis2023tomography}
For Ne interacting with H$_2^+ (1 ^2A^\prime)$
full dimensional electronic structure calculations at the current
``gold standard'' CCSD(T) (coupled cluster with single, double and
perturbative triples)\cite{bistoni:2020} using quintuple-zeta quality
basis sets have been carried out.~\cite{margulis2023tomography}. In contrast, full CI calculations are computationally too costly, but earlier PESs at the multi-reference configuration interaction (MRCI) are available\cite{lv2010exact,xiao2011quasi}. In the following we will compare three PESs which differ, in addition to the electronic structure method, in terms of utilised basis sets (quadruple vs quintuple) as well as sampled grid points and interpolation methods.

Experimentally, the dynamics on the ionic PES are initiated by Penning ionization~\cite{siska1993molecular} following collision of a
metastable Ne$^*$ with H$_2$. The collision on the neutral surface, prior to the Penning ionization, can be regarded as the ``first half'' of a process proceeding in two distinct steps.\cite{HensonScience2012,Tanteri2021} On the neutral surface, the initial state is a plane wave characterised by its energy. The second
``half-collision", on the ionic surface, starts with an initial wavepacket, with an approximately Gaussian shape, and a mean Ne--H$_2^+$ separation of $\sim 9\,\mathrm{a}_{0}$. Inbetween the two half-collisions, Penning ionization is modeled as a vertical transition to the electronic ground state of the molecular ion,\cite{ShagamNatChem15,margulis2023tomography} populating different vibrational states of Ne--H$_2^+(v,j)$. These decay to Ne + H$_2^+(v',j')$ and yield the
H$_2^+(v',j')$ translational kinetic energy spectrum as the main
observable. The pronounced angular anisotropy of the
Ne--H$_2^+$ interaction leads to rovibrational quenching for $v>0$
which converts vibrational into rotational and kinetic energy. 
This is reflected in distinct peaks in the H$_2^+$ kinetic energy spectrum~\cite{margulis2023tomography} which correspond to different
final $j-$rotational states. Each measured peak consists of
contributions from different total angular momenta $J \in \left\{|\ell-j|,\ldots,  \ell+j\right\}$ and partial
waves $\ell$~\cite{margulis2023tomography} and, most importantly,  several Feshbach resonances
which, unlike the final
rovibrational states, cannot easily be resolved in the experiment but
are amenable to computations. Such ``quantum fingerprints'' of the
collision dynamics are partially averaged out in the experiments due
to the finite energy resolution of the detector.

\begin{figure*}[tbp]
  \centering
\includegraphics[width=\textwidth]{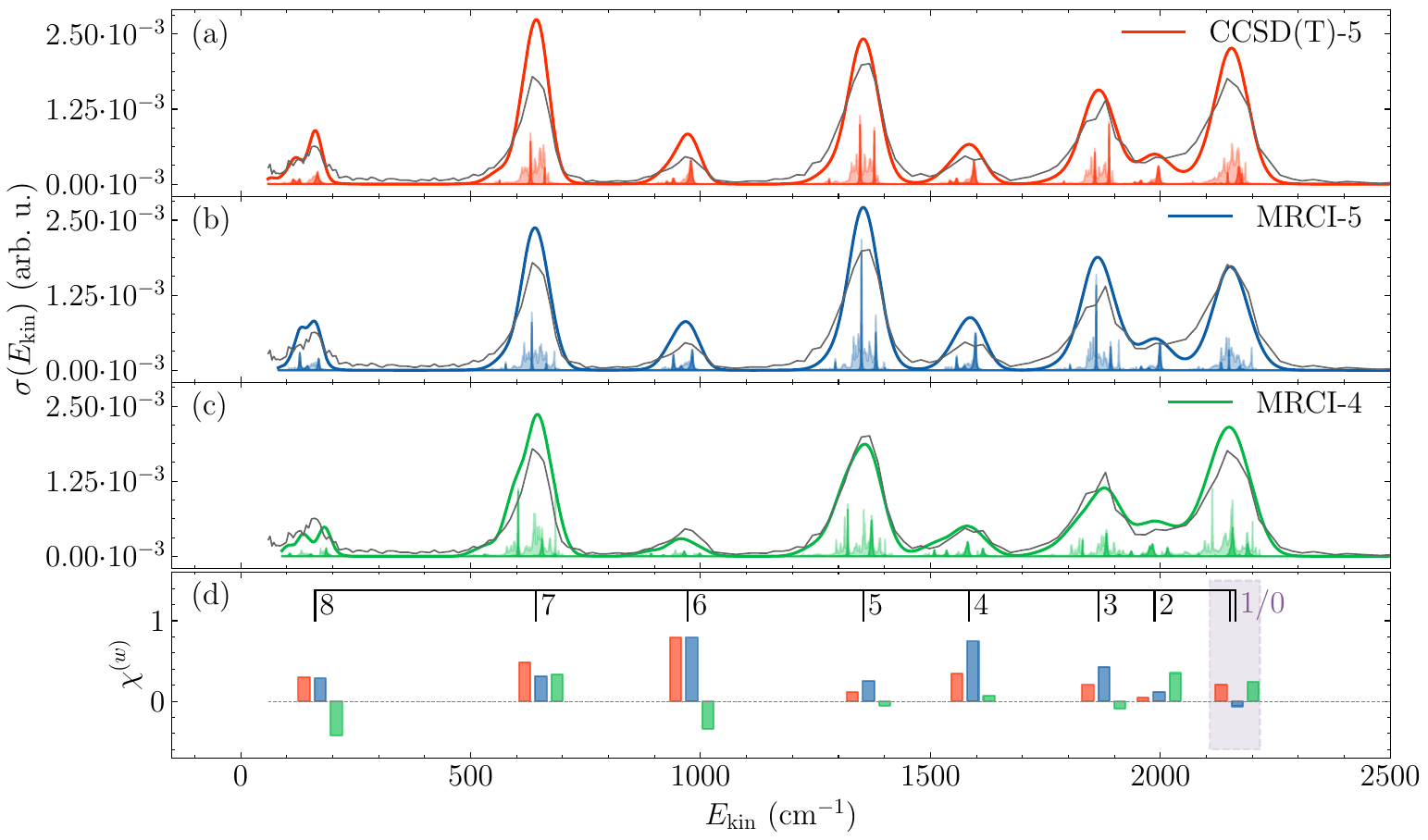}
\caption{a-c: Calculated cross sections (summed over all
  initial total angular momenta $J$, partial waves $\ell$ and
  product $v',j',\ell'$ and convoluted with the detector resolution)
  as a function of the kinetic energy, compared with the experimental
  data~\cite{margulis2023tomography} (grey) for an initial wavepacket
  in the diatomic rotational ground state and with $v=1$ 
  contrasted with
  its unconvoluted form (scaled to one tenth of its relative height,
  lightly shaded
  areas). Darkly shaded areas indicate the dominant total angular
  momentum and partial wave combinations $\ell=5$ and $J=5$ for both
  para- and ortho-H$_{2}^{+}$. d: Integrated deviations 
  $\chi^{(w)}$,
  cf. \cref{eq:Delta_peak},
  using the same color coding. The window highlighted in
  purple comprises two final $j'$ contributions as indicated by
  the comb. 
  }
    \label{fig:kinhistov1}
\end{figure*}

To assess the quality of the three PESs considered (CCSD(T)-5, MRCI-5,
MRCI-4; see SM), full coupled-channels quantum scattering calculations
were carried out.
  The physical system can be described using the three dimensional Jacobi coordinates $R$, $r$, and $\theta$, where $R$ describes the distance between Ne and the H$_2^+$ centre of mass, $r$ is the H$_2^+$ bond length and $\theta$ is the angle between the H$_2^+$ axis and the axis connecting Ne with the H$_2^+$ centre of mass.
The total energy of the system,  given as the sum of kinetic and internal energy, is conserved; its zero is chosen to correspond to the dissociation limit of the input channel, i.e., the channel of the initial wave packet, which  depends on $v$ and $j$. The kinetic energy is given in terms of the relative momentum of the two particles' center of mass frame $\hat{E}_{\mathrm{kin}}=\nabla^2_R/2\mu$.
The topography of the PESs
considered~\cite{margulis2023tomography,lv2010exact,xiao2011quasi} is
shown as supplemental material (SM)~\cite{SM}, with \ma significantly
deviating from \meuwly and \he in terms of minimal well-depth but also
angular anisotropy. In terms of long range behaviour, \meuwly agrees well with the behavior expected for the electrostatic interaction between a singly charged ion and a neutral particle,\cite{MM.heh2:2019} whereas MRCI is generally known to have difficulty in properly predicting the long range, which is also seen here for \he and \ma. In contrast, at short range, differences between \meuwly and \he are rather small. 
The calculation of the cross sections uses the computational framework detailed already in earlier work \cite{margulis2023tomography}.
Briefly, the total angular momentum $\vec J$ of the collision complex is conserved, as is parity. Here $\vec J=\vec L +\vec j$, where $\vec j$ describes molecular rotation and $\vec L$ the rotation around the center of mass of the collision complex.
Rovibrational quenching leads to changes in the partial wave quantum number $\ell$,  in order to preserve $J$ (with $J\in \left\{|\ell-j|,\ldots,
  \ell+j\right\}$). The basis for the coupled channels calculations is thus characterized by quantum numbers $J, M_J, j, \ell, v$~\cite{margulis2023tomography}, and each of the potentials is expanded in this basis. 
  
\begin{table*}[tbp]
\begin{NiceTabular}{|c  | c | c c c c |}
    \Hline
\Block{1-1}{$v$} &  PES & $\mathcal{F}$ &   $\mathcal{M}$ &$\Delta(0)$ & $E_{\mathrm{shift}}^{(\mathrm{opt})}$ \\
\Hline
\Block{3-1}{1}
& \meuwly\cite{margulis2023tomography}
& \colorbox{red!20!white}{$0.99$}& \colorbox{white}{$0.99$}& \colorbox{red!20!white}{$0.99$}& \colorbox{red!20!white}{$-0.74$}\\
& \he\cite{lv2010exact}
& \colorbox{white}{$1.2$}& \colorbox{white}{$1.0$}& \colorbox{white}{$1.6$}& \colorbox{white}{$21.0$}\\
& \ma\cite{xiao2011quasi}
& \colorbox{white}{$1.1$}& \colorbox{green!20!white}{$0.79$}& \colorbox{white}{$1.6$}& \colorbox{white}{$30.0$}\\
	\Hline
\Block{3-1}{2}
	& \meuwly& \colorbox{red!20!white}{$0.85$}& \colorbox{white}{$0.85$}& \colorbox{red!20!white}{$0.85$}& \colorbox{red!20!white}{$-0.54$}\\
	& \he& \colorbox{white}{$1.1$}& \colorbox{white}{$0.97$}& \colorbox{white}{$1.2$}& \colorbox{white}{$16.0$}\\
	& \ma& \colorbox{white}{$0.9$}& \colorbox{green!20!white}{$0.73$}& \colorbox{white}{$1.2$}& \colorbox{white}{$27.0$}\\
	\Hline
\end{NiceTabular}
\caption{
Agreement between calculated and experimental
  cross sections in terms of 
  $\mathcal F$,
  cf. Eq.~\eqref{eq:figure_of_merit}~\cite{SM}; the RMS deviation obtained for the
  energy shifted spectrum $\mathcal M$, 
  cf. Eq.~\eqref{eq:Delta_RMS}; the RMS
  deviation for the unshifted spectrum $\Delta(0)$; and the optimal
  energy shift $E_{\mathrm{shift}}^{\mathrm{(opt)}}$. Values are
  reported with normalization according to
  Eq.~\eqref{eq:cross_section_normalisation} and in multiples of $10^{-2}$ except for $E_{\mathrm{shift}}^{(\mathrm{opt})}$ in $\mathrm{cm}^{-1}$.  
  The PES performing best for a given
  quantifier is highlighted with the color code of
  \cref{fig:kinhistov1}. 
  }
\label{tab:results}
\end{table*}

\Cref{fig:kinhistov1} shows the translational kinetic energy spectra obtained with the three
PESs for initial wavepackets with $v=1$ and convoluted with the
experimental resolution. The spectra are shifted along the kinetic
energy axis to minimize the root-mean-square (RMS) difference between
computed and observed peak positions. Analogous data for $v=2$ is
shown as SM~\cite{SM}. Without shifting, the peak positions can differ
by up to $30\wavenum$, see \cref{tab:results}. At this stage, the
quality of the CCSD(T) PES is superior to that of the two MRCI-based
PESs. However, it should also be noted that \he and \ma are reactive
PESs and were developed to investigate proton transfer between H$_2^+$
and Ne. The peak positions quantify the amount of internal energy that
was converted into kinetic energy and thus directly reflect the energy
of the FRs. Once this source of deviation between prediction and
measurement is accounted for, the remaining comparison focusses on the
distribution of the converted energy over the different final
rovibrational states, see \cref{fig:kinhistov1}(a)-(c). The latter is
mainly determined by the angular anisotropy. Given the experimental
resolution, all three PESs compare reasonably well with the
experimental data. The broad peaks in \cref{fig:kinhistov1}(a)-(c),
most of them well separated in $v'$ and $j'$, correspond to the
different \textit{final} state contributions, i.e. it is almost always
possible to resolve the final H$_2^+ (v',j')$ states with the current
experimental resolution~\cite{margulis2023tomography}. In contrast,
while without convolution most of the peaks in the kinetic energy
spectra can be attributed to a specific \textit{initial} Feshbach resonance state (with
$J,\ell,j$, shown in light shade), this information is lost after
convolution. The bare, i.e., unconvoluted cross sections display much
larger differences between the PESs.

To quantify the agreement for the peak positions, we introduce the
energy $E_{\mathrm{shift}}^{(\mathrm{opt})}$ by which the spectra have
to be shifted to optimally match the experimental peak positions.
This shift has to be compared to the (velocity-dependent) experimental
uncertainty in the peak positions which is $\sim 20\,
\mathrm{cm}^{-1}$ for the low-$j'$ peaks but only $2\wavenum$ for the
high-$j'$ peaks. The smallest shift by far is needed for the cross
sections obtained with the \meuwly PES, cf. \cref{tab:results}. To
quantify the matching in terms of the peak heights, the (signed)
differences $\chi^{(w)}$ are evaluated for an energy window $w$ to
assess the agreement on a per-peak basis, see
\cref{fig:kinhistov1}(d). This confirms the comparable performance of
all three PES in terms of the peak heights, in particular for the
strong peaks. Finally, a single figure of merit $\mathcal F$ was
obtained by averaging the RMS difference with respect to the energy
shift up to the optimal value $E_{\mathrm{shift}}^{(\mathrm{opt})}$,
see SM. This figure of merit is designed to reward PESs that correctly predict the energies at which the Feshbach resonances occur whilst also ensuring that the distribution of peak intensities is correct. In terms of $\mathcal{F}$, the \meuwly PES yields the closest match with the
experimental data for both $v=1$ and 2, cf. \cref{tab:results}. This
is mainly due to the small energy shift
$E_{\mathrm{shift}}^{\mathrm{(opt)}} \leq 1$ cm$^{-1}$ required.


\begin{figure*}[tbp]
    \centering
    \begin{tikzpicture}
        \node (top) at (0,0) {\includegraphics[width=\textwidth]{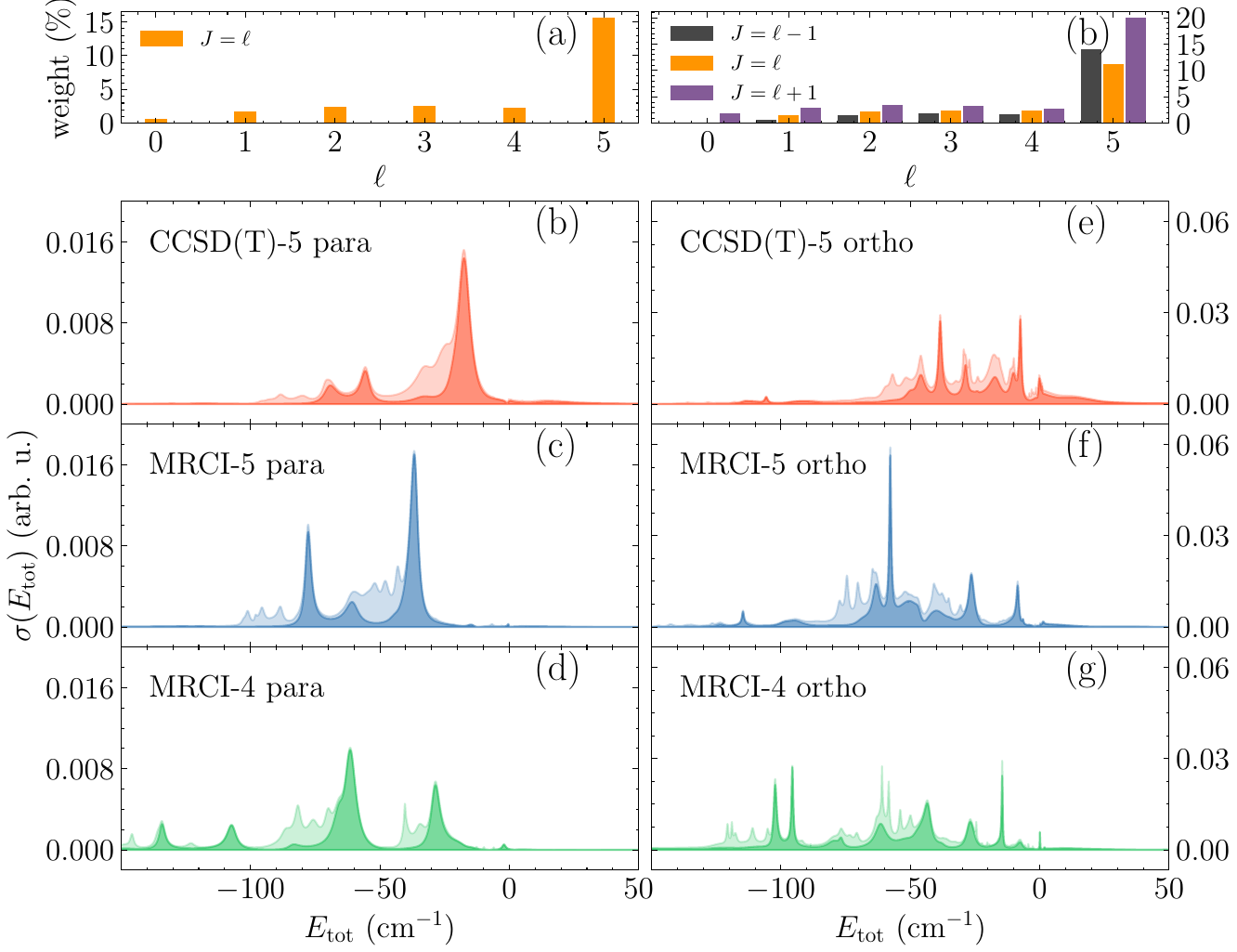}};     
    \end{tikzpicture}
    \caption{Calculated collision cross sections (b-d, e-g) as a function of the
      total energy. The top set of panels (a,b) shows the weighting of initial total angular momentum $J$ and partial wave $\ell$ contributions. 
      In the bottom set of panels, (b-d) correspond 
      to $v=1$ and para-H$_2^+$ with
      initial $j=0$, whilst (e-g) correspond to ortho-H$_2^+$ with initial $j=1$.
      The lightly shaded backgrounds denote the total spectra, whilst the specific contributions due to the (dominant) initial $\ell=5$ partial wave are portrayed as darkly shaded slabs.}
    \label{fig:cross_section_collision_energy}
\end{figure*}

The advantage of the \meuwly PES derives from the fact that the RKHS
representation of the CCSD(T) energies results in more accurate
positions of the Feshbach resonances. 
Focusing on the cross section as a function of the total energy allows
for a more in-depth analysis of the Feshbach resonances,
cf. \cref{fig:cross_section_collision_energy}. As a function of total
energy, both peak shapes and positions of the cross sections differ
vastly between the PES. In particular, a clear bias towards lower
energies is seen for \he and \ma as compared to \meuwly. This is true for both $v=1$
and $v=2$ (shown in SM~\cite{SM}) and para- as well as ortho-H$_2^+$. For para-H$_2^+$ the cross sections for both $v=1$ and
$v=2$ are primarily due to $\ell=5$, $J=5$ (shown with dark shade) and
are comprised of three main peaks, corresponding to well-isolated
FRs. In contrast, for ortho-H$_2^+$, where the dominant partial wave
contribution consists of three different total angular momenta $J$,
the cross sections indicate several, partially overlapping FRs,
  cf. \cref{fig:cross_section_collision_energy} (e)-(g) for $v=1$ (the data for $v=2$ is shown as SM in \cref{fig:v2_cross_section_collision_energy} (d)-(f)).
Focusing on $v=1$ for para-H$_2^+$,
\cref{fig:cross_section_collision_energy} (a)-(c), our analysis is
facilitated by the fact that a single resonance, around
$E=-17~\mathrm{cm}^{-1}$, dominates for \meuwly whilst the cross
sections obtained with \he and \ma are both comprised of two
significant contributions, occurring at lower energies, around
$-30\wavenum$ and $-70\wavenum$, respectively.

\begin{figure}[tb]
    \includegraphics[width=\linewidth]{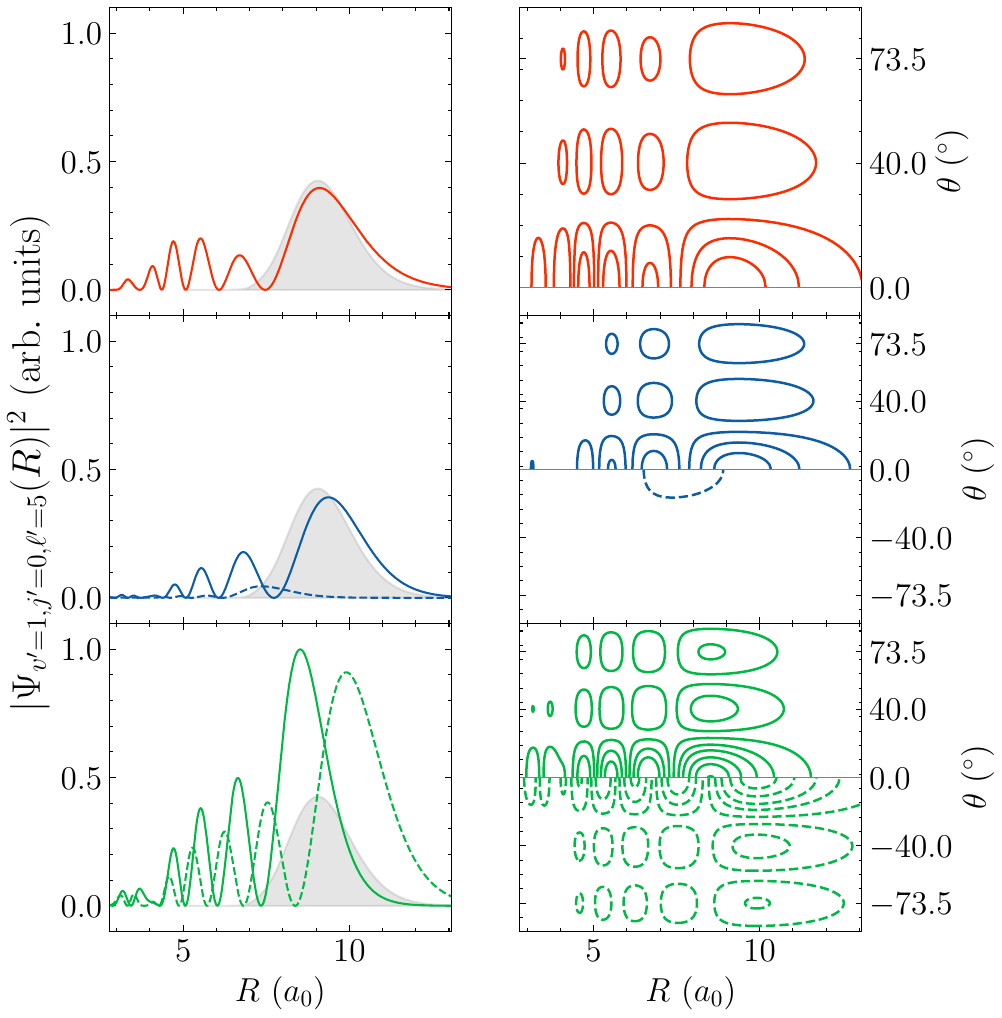}
    \caption{Components of the resonance wavefunctions' \change{squared amplitude} with
    $v'=1$, $j'=0$ and $\ell'=5$, cf.  \cref{fig:cross_section_collision_energy} (a-c). 
    The resonance energies are $-17.45~\mathrm{cm}^{-1}$ for \meuwly in red, $-36.75~\mathrm{cm}^{-1}$ (solid) and $-77.65~\mathrm{cm}^{-1}$ (dotted) for \he in blue  and $-61.45~\mathrm{cm}^{-1}$ (solid) and $-28.45~\mathrm{cm}^{-1}$ (dotted) for \ma in green.
    Where applicable, the two highest resonances are shown in the positive and negative angular domains of the contour plots, respectively.
    The shaded grey areas show the shape of the initial wave function on the ionic surface.}
    \label{fig:resonance_wfs}
\end{figure}
The cross section peaks in \cref{fig:cross_section_collision_energy} are linked
to the topology of the PES and the shapes of the resonance
wavefunctions. At small $R$ the \ma PES differs from the \meuwly and
\he PESs, especially by its much deeper well.\cite{SM} 
The \change{squared amplitude} of the most important resonance wavefunctions are shown in \cref{fig:resonance_wfs}, where the left set of panels also compares them with the "initial" wave packet (grey shaded curve), i.e.,  $v=v'=1$, $j=j'=0$ and $\ell=\ell'=5$. 
The wavefunction amplitude covers a large range of inter-particle distances, from the strongly interacting region
($R\approx3.4\,$a$_0$ to $3.8\,$a$_0$, depending on the PES) where the anisotropy is most pronounced, all the way to long range interactions.
Here it becomes apparent that the (main) resonances for \meuwly and \he show significant overlap with the input wavepacket. In contrast, for \ma, the two main resonances are centered around slightly shorter, respectively larger distances as compared to the input wavepacket. They also have much larger amplitudes relative to the other two surfaces. The right set of panels in \cref{fig:resonance_wfs} demonstrates the angular behaviour of these resonance wave function components, by taking the partial wave quantum number into account. 
The similar topologies of the \meuwly and \he PES at short range correspond with their peak intensities matching each other more closely than those obtained with the \ma PES, even though the energies at which these resonances occur are appreciably different, c.f. \cref{tab:results}. In contrast, for large $R$ the \meuwly PES differs most, whilst the two MRCI PESs show similar long range behavior. Given the best performance of \meuwly in terms of
$E_{\mathrm{shift}}^{\mathrm{(opt)}}$, combined with the fact that the two MRCI PESs demonstrate both similar long-range behavior as well as comparable values for $E_{\mathrm{shift}}^{\mathrm{(opt)}}$, suggests that the long range
behavior of the PES primarily determines the peak positions. This
confirms earlier findings from quantum wavepacket simulations for
half-collisions between He and H$_2^+$ molecules~\cite{horn2023improving}.

The conspicuously substantial long-range anisotropy of the \ma PES and the
resulting difference in wave functions and cross sections indicate that the peak heights in the final $v', j'$-distribution are particularly sensitive to the potential at short and intermediate $R$. This is where many avoided crossings between the adiabatic potential energy curves for each $v,j$
(resulting from an adiabatic separation of vibrational and rotational
motion) are observed~\cite{margulis2023tomography}. Passage through
the crossings redistributes the energy to the various final $v', j'$
states which in turn is reflected in the peak height of the cross
sections.

\begin{figure}[tb]
    \includegraphics[width=\linewidth]{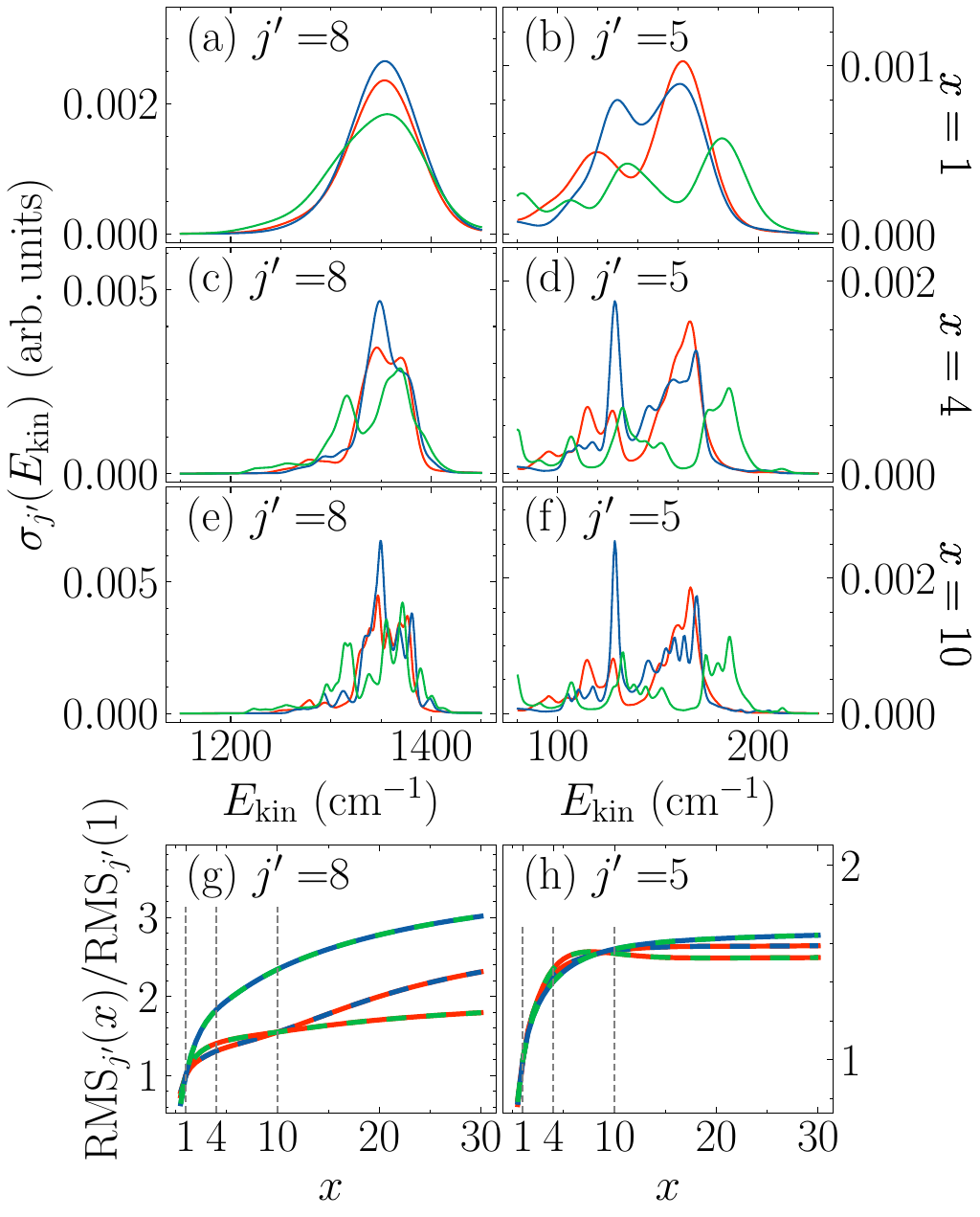}
    \caption{Comparison of the cross sections 
      calculated with three different PES for increasing
      experimental resolution $x$ and 
      $j'=8$ (a,c,e) and $j'=5$ (b,d,f) and 
      change in
      RMS deviation for each pair of PES (indicated by the
      color coding) as the resolution is enhanced (g,h).  
      Same color for PES as before.
       }
    \label{fig:rms_resolution_scaling}
\end{figure}

The figure of merit $\mathcal{F}$ for the \meuwly PES is lower by 5 \%
to 20 \%, respectively, compared with the \he and \ma PESs, depending
on the vibrational state $v$ considered, see Table
\ref{tab:results}. Based on this, we now address the question by how
much the resolution of the experiments needs to be improved in order to resolve
individual $\ell-$components to further validate the PESs. To this
end, kinetic energy spectra for three different experimental
resolutions -- the current value as well as a four-fold, resp.~ten-fold improved resolution -- are compared in
\cref{fig:rms_resolution_scaling}. For better visibility, we focus on 
one exemplary peak, 
$j'=8$ for ortho-H$_2$ in the right and $j'=5$ for para-H$_2$ in the left part  of \cref{fig:rms_resolution_scaling}.  When scaling the
resolution, the convolution width is assumed to be independent of the
kinetic energy whereas a kinetic energy-dependent convolution width
accounts for experimental uncertainty~\cite{margulis2023tomography}.
\Cref{fig:rms_resolution_scaling} shows the convoluted final state
distribution at the present experimental
resolution~\cite{margulis2023tomography} 
with fixed convolution widths (as opposed to being energy-dependent) for better comparability, along
with the curves for four-fold and ten-fold improved resolutions (with
fixed widths). The differences between the fixed-width and
energy-dependent convolution widths are minor.

At four times the current experimental resolution, the $\ell$
substates can be identified and the convoluted spectra display a
different energy dependence for the different PESs. For the
contribution of the dominant initial partial wave ($\ell=5$) and in
some cases also for the initial partial waves with lower weight,
individual peaks in the spectra can be attributed to specific FRs
occurring at different energies, for example the peaks at kinetic
energies of $175\wavenum$ and $625\wavenum$. 
For example, in the case of \meuwly,
three distinct peaks, seen at roughly $115\wavenum$, $130\wavenum$ and
$170\wavenum$ in \cref{fig:rms_resolution_scaling}(d); they correspond to
the three peaks at $-70\wavenum$, $-55\wavenum$ and $-20\wavenum$ in
\cref{fig:cross_section_collision_energy}(a).  For \he, two pronounced
peaks are seen at $130\wavenum$ and $170\wavenum$ and a flatter peak
at roughly $140\wavenum$ in \cref{fig:rms_resolution_scaling}(d), which
can be attributed to the various peaks occurring at about
$-80\wavenum$, $-60\wavenum$ and $-35\wavenum$ in
\cref{fig:cross_section_collision_energy}(b).  \ma, in contrast,
displays broad peaks at around $110\wavenum$, $150\wavenum$ and
$185\wavenum$ in \cref{fig:rms_resolution_scaling}, corresponding to the
$-110\wavenum$, $-60\wavenum$ and $-30\wavenum$ peaks in
\cref{fig:cross_section_collision_energy}(c).  To conclude, enhancing
the energy resolution by a factor of four has a two-fold effect: On
top of deciding which of the PES best captures the details of the
interaction, it will also allow for assigning the peaks in the kinetic
energy spectrum to specific FRs.

At ten times the current experimental resolution, the splitting is
much more pronounced and several peaks can be clearly attributed to
different initial $J,\ell$ channels. Most importantly, however, the collision complex features sufficiently many,
energetically well-isolated FRs such that the shapes of the convoluted
cross sections differ in a pronounced fashion amongst the three PES. This is discussed in more detail in the SM\cite{SM}.

The largest gain in improving the ability to
differentiate the PESs is observed when increasing the resolution by a
factor of four. In order to make this observation more quantitative,
\cref{fig:rms_resolution_scaling}(g,h) shows the peak-wise RMS
deviation between pairs of simulated spectra (indicated by the color
code) integrated over the respective energy windows shown in
\cref{fig:rms_resolution_scaling}(a-f).  A large
increase in the RMS deviation corresponds to an increase in the
ability to distinguish two theoretical predictions from each other,
and thus also their respective comparison with the experimental
data. As one would expect, increasing the resolution will not lead to
an improved distinguishability indefinitely. At which resolution
saturation sets in depends on the specific peak, i.e., the final $j'$
values: While no substantial further improvement is observed in
\cref{fig:rms_resolution_scaling}(h) when increasing the energy
resolution by more than a factor of ten, in \cref{fig:rms_resolution_scaling}(g) the distinguishability continues
to increase gradually even until 30 times the original resolution.

While a ten-fold increase in the kinetic energy resolution compared to
the recent experiment~\cite{margulis2023tomography} may prove very
challenging, a four-fold increase will require only moderate changes
to the existing setup. With the corresponding kinetic energy
resolution, it will already be possible to attribute a good part of
the kinetic energy spectrum to specific initial \textit{and} final
states, taking the experiment a big step towards fully resolved
`quantum tomography' of the collision. This will come on top of the
ability of the measurement to decide which level of theory for the
interparticle interactions captures physical reality best.

\begin{acknowledgement}
We acknowledge financial support from the Swiss National Science
Foundation (NCCR MUST, 200020\_219779, 200021\_215088), the AFOSR (award number FA8655-21-1-7048), and the University of Basel (to MM).
\end{acknowledgement}

\textbf{Supporting Information Available:} Visualization of PES, details of the figures of merit for comparison, calculated cross sections for an initial wavepacket with $v=2$, scaling of the convolution resolution for additional peaks and breakdown of individual partial wave contributions.

\bibliography{refs}

\newpage

\section{Supplementary Material}

\newpage

\newcounter{Sequ}
\newenvironment{SEquation}
  {\stepcounter{Sequ}%
    \addtocounter{equation}{-1}%
    \renewcommand\theequation{S\arabic{Sequ}}\equation}
  {\endequation}

\newenvironment{SEqnarray}
    {\stepcounter{Sequ}
    \addtocounter{equation}{-1}
    \renewcommand\theequation{S\arabic{Sequ}}\eqnarray}    
    {\endeqnarray}

\newenvironment{SAlign}
    {\stepcounter{Sequ}
    \addtocounter{equation}{-1}
    \renewcommand\theequation{S\arabic{Sequ}}\align}
    {\endalign}

\renewcommand{\thetable}{S\arabic{table}}
\renewcommand{\thefigure}{S\arabic{figure}}
\setcounter{figure}{0}

\section{Potential energy surfaces}
The three PESs differ in the underlying
  approximations involved in the electronic structure methods (CCSD(T)
  vs MRCI), the basis sets used (aug-cc-pV5Z, indicated by the suffix
  `-5', and aug-cc-pVQZ, indicated by the suffix `-4' in the
  following), and their representations (reproducing kernel Hilbert
  space vs. parametrized fits).
The CCSD(T)-5 PES is a reproducing kernel Hilbert 
space~\cite{unke2017toolkit} (RKHS) representation of CCSD(T)/aug-cc-pV5Z 
reference energies computed using MOLPRO~\cite{werner2020molpro} for a grid of 
Jacobi coordinates ($R, r, \theta$). Here, $r$ is the H$_2^+$
diatomic bond length, $R$ is the distance between the center of mass of the diatomic 
molecule and the neon atom and $\theta$ is the angle between $\vec{r}$ and $\vec{R}$.
The grid (on-grid points) included 39 points for $r \in [1.1\,\mathrm{a}_0,8.0\,\mathrm{a}_0]$, 49 points for $R \in [1.0\,\mathrm{a}_0, 
45.0\,\mathrm{a}_0]$ and 10 Gauss-Legendre quadrature points for $\theta \in [0^\circ, 90^{\circ}]$.\cite{margulis2023tomography}
The root mean squared errors (RMSE) for energies of 250 geometries that 
were not used in constructing the RKHS representation (off-grid points) 
between reference calculations and the evaluated RKHS is 11.8 
cm$^{-1}$, compared with an RMSE of $11.3$ cm$^{-1}$ for on-grid 
points.~\cite{margulis2023tomography} This compares with RMSEs of 94 cm$^{-1}$ (0.27 kcal/mol)  and 315 cm$^{-1}$ (0.0391 eV) for the MRCI-5 and MRCI-4 PESs, respectively.\cite{lv2010exact,xiao2011quasi}

\begin{figure*}
    \includegraphics[width=\textwidth]{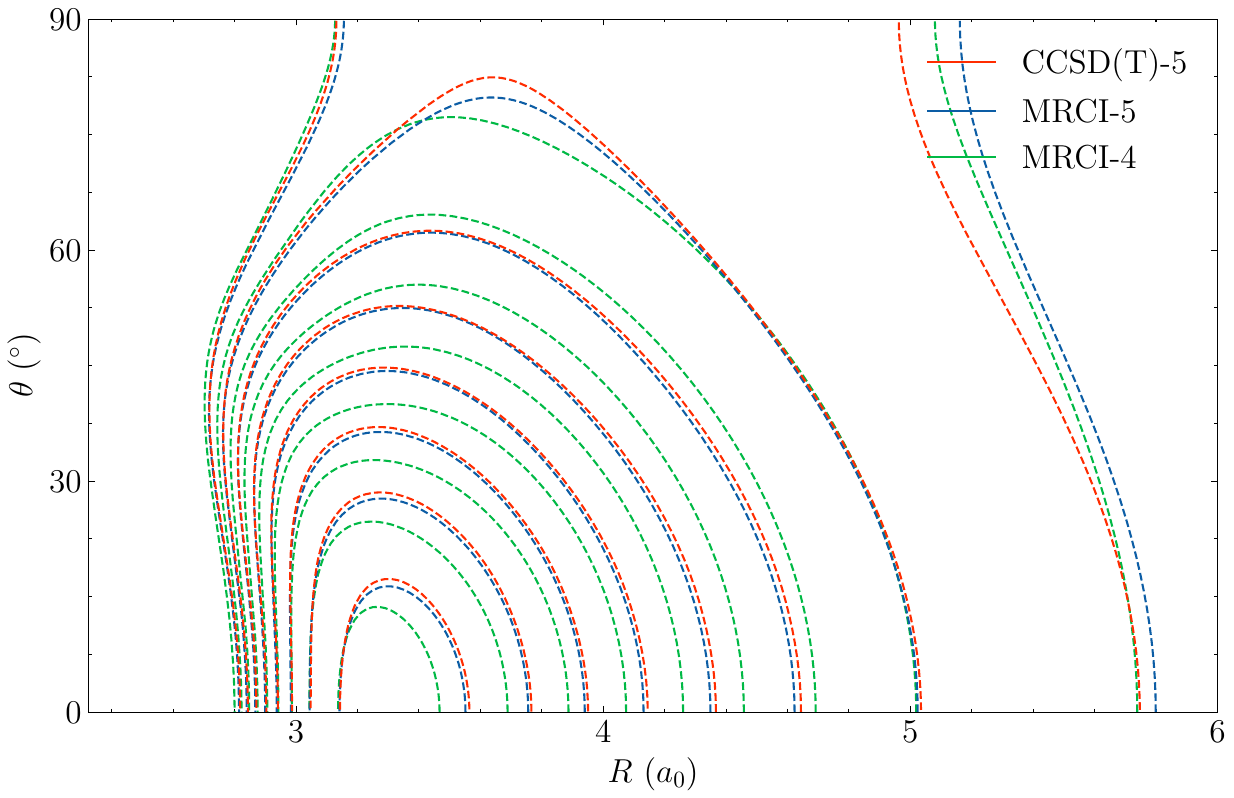}
    \caption{Comparison of potential energy surfaces for a specific slice of coordinate space ($r=2.0\,a_0$) in the strongly interacting, bound state region.
    Each potential energy surface has the same set of contours (starting at $-500 \,\mathrm{cm}^{-1}$ around $R\approx 5.8~a_0$ and decreasing in steps of $500 \,\mathrm{cm}^{-1}$ towards the minimum at $R\approx 3.3\,a_0$ (note that the \ma surface has an additional contour corresponding to $-4500\,\mathrm{cm}^{-1}$, which is lower than the minimal energies of \meuwly and \he for the shown slice). The zero of energy was determined as the potential energy surface value at $r=2.0\,a_{0}$ and $R=1000\,a_0$. 
    }
    \label{sfig:2dpes}
\end{figure*}

\begin{figure*}
    \centering
    \includegraphics[width=0.9\linewidth]{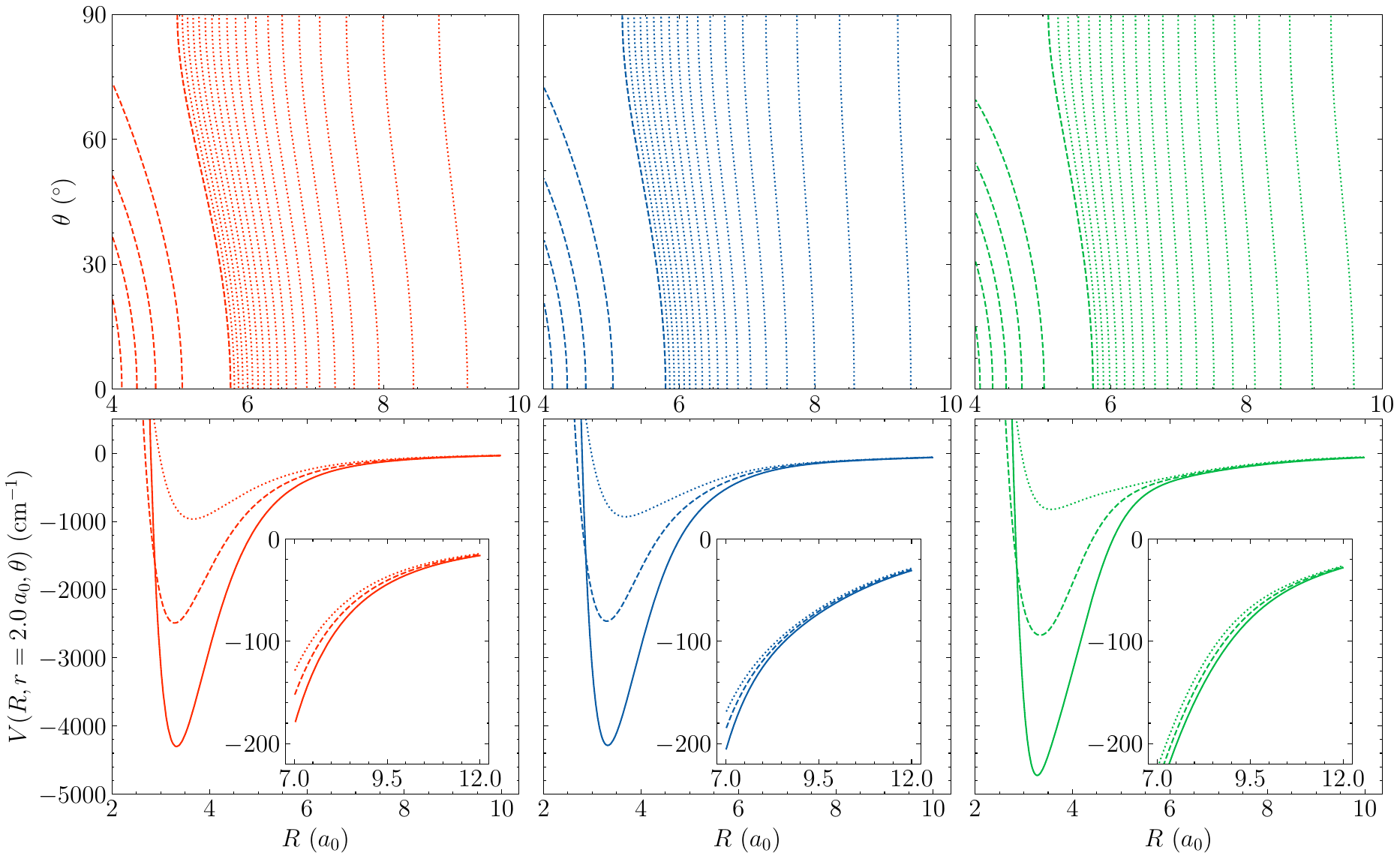}
    \caption{Comparison of PESs focusing on the long-range part. Top panels from left to right: PESs \meuwly , \he and \ma for $R \in [4,10]$ a$_0$ and $r=2.0\,a_{0}$. With increasing $R$ the anisotropy for \he reduces considerably compared with the other two PESs. Spacing between dashed line contours is 500 cm$^{-1}$ as before whilst the spacing between dotted contours is 25 cm$^{-1}$. 
     Bottom panels: selected one dimensional cuts for $\theta=0^{\circ}$ (solid), $45^{\circ}$ (dashed) and $90^{\circ}$ (dotted lines), with $r=2.0\,a_{0}$.
     Inset panels show a close up of the long range behaviour, emphasising the differences in both spread and anisotropy even at large $R$. As above, the \meuwly , \he and \ma contours and graphs are shown in red, blue and green respectively.}
    \label{fig:PES-comparison}
\end{figure*}

\Cref{sfig:2dpes} shows a two dimensional slice in $\theta$ and $R$
of the various potential energy surfaces at $r=2.0~a_0$,
which is near to the H$_2^+$ equilibrium distance.
At short range there is close agreement between \meuwly and \he for all angles 
$\theta$, whilst \ma shows large deviations (both radially and angularly)
from the other two that grow larger in the vicinity of the minimum around 
$R\approx 3.4~\mathrm{a}_0$.

\begin{figure*}
    \includegraphics[width=\textwidth]{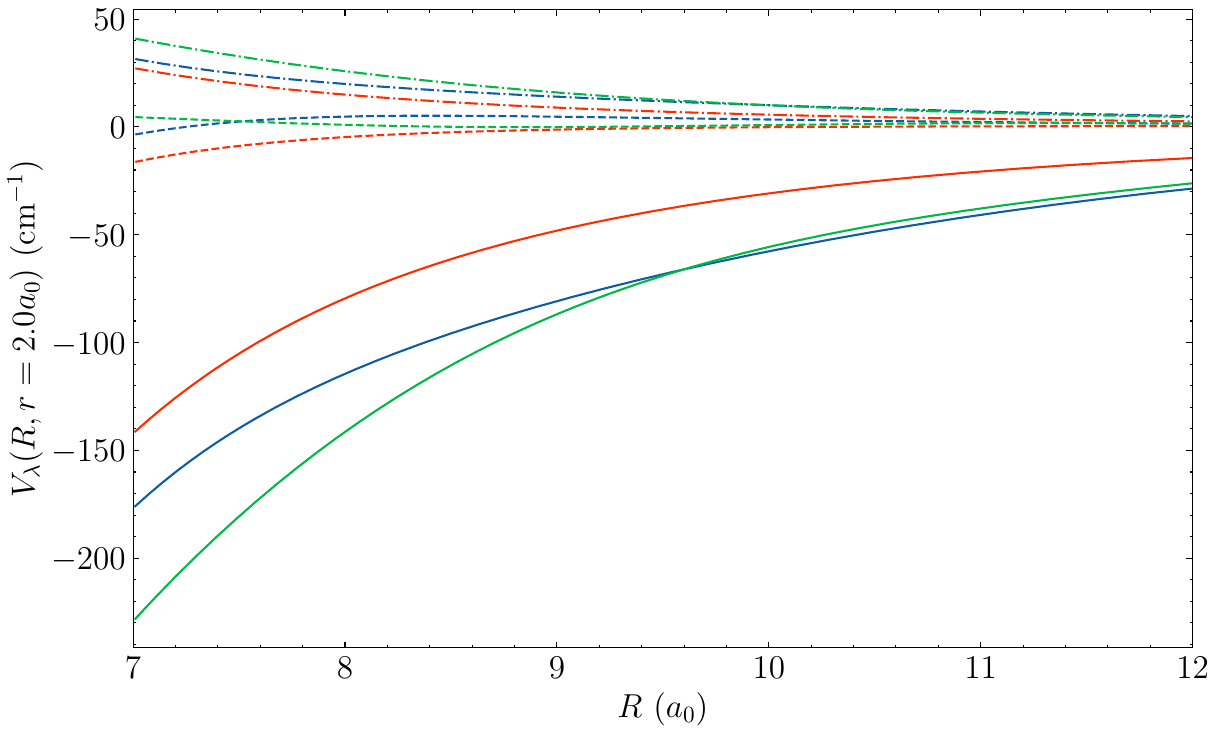}
    \caption{Comparison of the long range behaviour of the various potential energy surfaces, portrayed in terms of
    the $R$-dependent Legendre coefficients $V_{\lambda}(R,r)$ for a fixed equilibrium distance $r=2~\mathrm{a}_{0}$.
    Large deviations are seen between the isotropic $\lambda=0$ curve for \meuwly and those
    corresponding to the other two potential energy surfaces.
    The solid, dashed and dash-dotted lines correspond to $\lambda=0,2$ and $4$ respectively.
    }
    \label{sfig:long_range_pes}
\end{figure*}

In contrast, \cref{fig:PES-comparison} and
\cref{sfig:long_range_pes} demonstrate the long
range behaviour of the potential in terms of the Legendre
expansion coefficients $V(R,r,\theta) = \sum_{\lambda=0}^{\infty} V_{\lambda}(R,r)P_{\lambda}(\cos\theta)$ at $r=2.0~a_{0}$, which is near to the 
equilibrium distance of H$_2^{+}$.
We observe that as the interparticle distance $R$ increases,
the isotropic term $\lambda=0$ of the \meuwly potential energy
surface shows a large deviation from both of the other potential 
energy surfaces.

\section{Figures of merit}
\label{sec:methods}

To compare cross sections with each other, the theoretical and experimental cross sections 
shown above have been `normalised' according to
\begin{SEquation}
    a \int_{\mathbb{R}_{+}} dE_{\mathrm{kin}} \tilde{\sigma}_{\mathrm{exp}} (E_{\mathrm{kin}})=
    b \int_{\mathbb{R}_{+}} dE_{\mathrm{kin}} \tilde{\sigma}_{\mathrm{theory}} (E_{\mathrm{kin}})=1
    \label{eq:cross_section_normalisation}
\end{SEquation}
for each potential energy surface.
In \cref{eq:cross_section_normalisation}, $\tilde{\sigma}_{\mathrm{exp}}(E_{\mathrm{kin}})$ and $\tilde{\sigma}_{\mathrm{theory}}(E_{\mathrm{kin}})$
are the cross sections prior to normalisation, whilst their rescaled counterparts appear without tildes.
We discuss potential drawbacks of this scaling method and introduce an alternate method below. In particular, we show how this leads to a better overall agreement
in terms of RMS deviation, but at the cost of an increase in the complexity.

When quantifying how well the calculated cross sections match the experimental cross section,  we separate the mismatch in Feshbach resonance energies from the mismatch in the final state distribution, i.e., the amplitudes of the peaks corresponding to a given $v'$ and $j'$.
For each potential energy surface we find the most favourable
$E_{\mathrm{shift}}$ by solving the minimization problem, 
\begin{SAlign}
    \label{eq:Delta_RMS}
    \min_{E_{\mathrm{shift}}}& \Delta(E_{\mathrm{shift}}) \\
    =& \min_{E_{\mathrm{shift}}} \sqrt{\int_{\mathbb{R}_{+}}
        dE_{\mathrm{kin}} 
    \left|\sigma_{\mathrm{exp}}(E_{\mathrm{kin}}) - \sigma_{E_{\mathrm{shift}}}(E_{\mathrm{kin}}) \right|^2}\,\,, \nonumber
\end{SAlign}
with 
  $$  \sigma_{E_{\mathrm{shift}}}(E_{\mathrm{kin}}) = \sigma_{\mathrm{theory}}(E_{\mathrm{kin}} + E_{\mathrm{shift}})\,
$$
the energy shifted cross section, 
using the NLopt python package~\cite{nlopt}.
$\Delta(E_{\mathrm{shift}})$ in \cref{eq:Delta_RMS},
is the root-mean-square difference between the (energy-shifted)
theoretical and experimental cross sections, which quantifies the
error in the distribution of peak heights and shapes. 
We will refer to the optimal shift (in terms of the best match of peak positions) as $E_{\mathrm{shift}}^{(\mathrm{opt})}$.

Neither $\Delta(E_{\mathrm{shift}}=0)$, which emphasises the correctness of the Feshbach resonance energies over the correctness of peak amplitudes, nor $\Delta(E_{\mathrm{shift}}^{\mathrm{(opt)}})$,
which emphasises the correctness of the peak amplitudes over the peak positions, alone can fully characterize the agreement between calculated and experimental cross sections. For a fair comparison, we combine $E_{\mathrm{shift}}^{(\mathrm{opt})}$ and
the root-mean-square deviation into a single figure of merit.
To this end, we average  $\Delta(E_{\mathrm{shift}})$
over the interval $\left[\min(0,E_{\mathrm{shift}}^{(\mathrm{opt})}),\max(0,E_{\mathrm{shift}}^{(\mathrm{opt})})\right]$, i.e.,
\begin{SAlign}
    \label{eq:figure_of_merit}
    \mathcal{F} = \frac{\int_{0}^{E_{\mathrm{shift}}^{(\mathrm{opt})}} dE_{\mathrm{shift}}\Delta(E_{\mathrm{shift}})}{\int_{0}^{E_{\mathrm{shift}}^{(\mathrm{opt})}} dE_{\mathrm{shift}}}\,\,.
\end{SAlign}
For a more fine-grained quantification of the differences,
it is instructive to divide up kinetic energy histograms 
into windows $w = [w_{\min},w_{\max}]$ corresponding to distinct 
peaks, with $\mathcal{W}=\left\{w\right\}$, the set of all windows.
Unsigned integral deviations for matching theory 
and experiment on a per peak basis are given by~\cite{margulis2023tomography}
\begin{SEquation}\label{eq:Delta_peak}
    \chi^{(w)}=\frac{\int_{w_{\min}}^{w_{\max}} d E_{\mathrm{kin}} \left(\sigma_{E_{\mathrm{shift}}}(E_{\mathrm{kin}}) -\sigma_{\mathrm{exp}}(E_{\mathrm{kin}}) \right)}
    {\int_{w_{\min}}^{w_{\max}} d E_{\mathrm{kin}} \sigma_{\mathrm{exp}}(E_{\mathrm{kin}})}\,\,.
\end{SEquation}
The deviations in \cref{eq:Delta_peak} take into account the shapes of the individual peaks and not just their maximum intensities.

\section{Cross section for initial vibrational excitation $v=2$}
\label{ssec:v2}

\begin{figure*}[tbp]
  \centering \includegraphics[width=\textwidth]{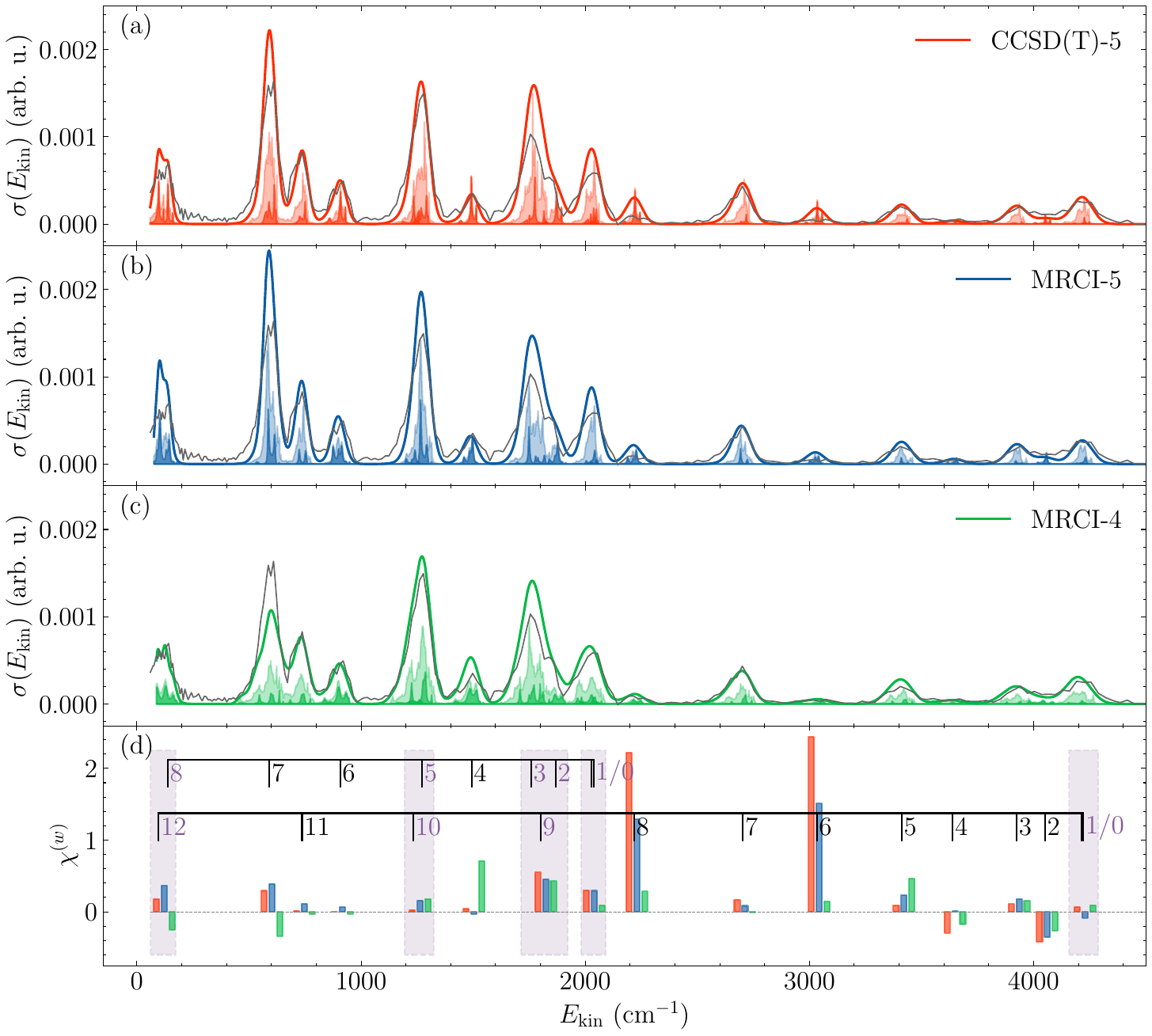}
    \caption{Same as \cref{fig:kinhistov1} but for the initial
      wavepacket in the diatomic rotational ground state and with $v=2$. The unconvoluted cross sections have  
      been scaled by a factor of three tenths relative to their
      convoluted counterparts.  }
    \label{fig:kinhistov2}
\end{figure*}
\Cref{fig:kinhistov2} shows the kinetic energy spectra
obtained with the three PESs, for initial wavepackets with $v=2$, convoluted with the experimental resolution, and shifted along the kinetic energy axis to minimize the root-mean-square (RMS) difference between computed and observed peak positions. It complements the data for $v=1$ shown in \cref{fig:kinhistov1}  in the main text. Note that for $v=2$ differences between the three PES are sufficiently large to remain visible even after
convolution, most noticeably so for the peak around $600\wavenum$ in \cref{fig:kinhistov2} corresponding to the $v=2,j=1
\rightarrow v'=1,j'=7$ transition.

\begin{figure*}[tbp]
    \includegraphics[width=\textwidth]{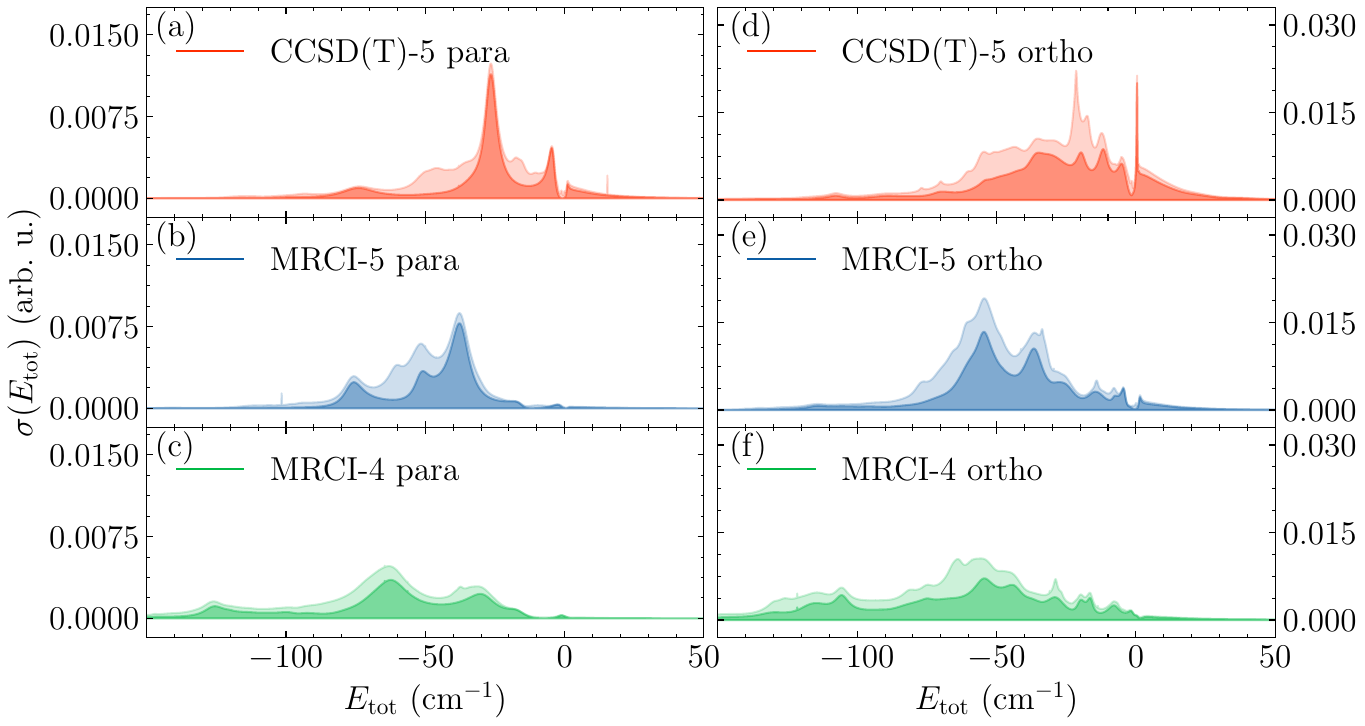}
    \caption{Same as \cref{fig:cross_section_collision_energy}, but for $v=2$.}
    \label{fig:v2_cross_section_collision_energy}
\end{figure*}

Similarly, \cref{fig:v2_cross_section_collision_energy} shows the cross section as a function of the total energy for a double initial vibrational excitation, in analogy to \cref{fig:cross_section_collision_energy} for $v=1$.
The $v=2$ cross section shows the same variation in FR structure observed in $v=1$ but in contrast to a single vibrational excitation, most FRs appear as broader, less pronounced peaks.

\section{Optimized scaling}
\label{ssec:optimised_scaling}

\begin{figure*}
    \includegraphics[width=\textwidth]{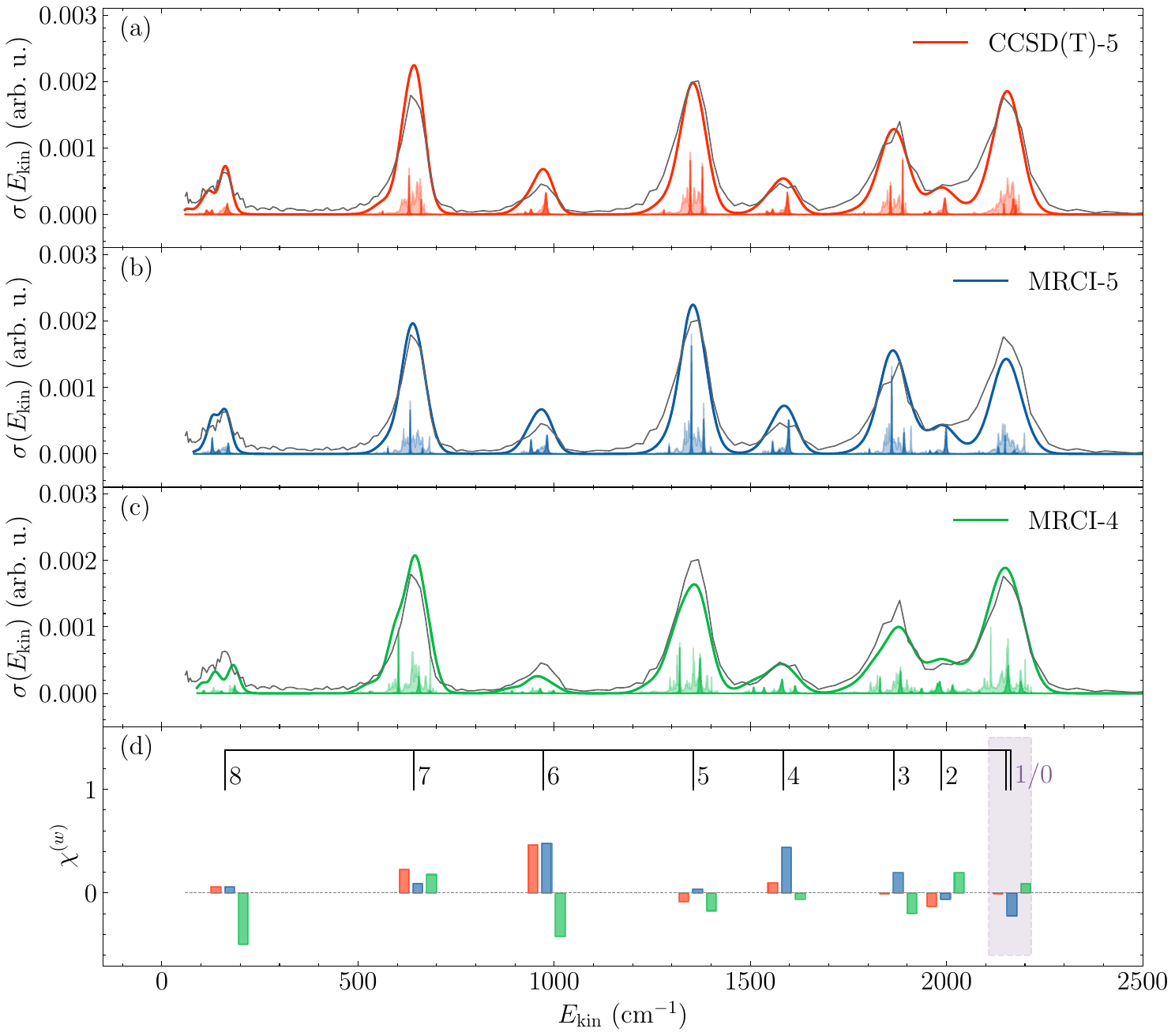}
    \caption{Same as \cref{fig:kinhistov1}, but with the fitted relative scaling
    of \cref{seq:Delta_RMS}.
    }
    \label{sfig:opt_scale_kinhistov1}
\end{figure*}

\begin{figure*}
    \includegraphics[width=\textwidth]{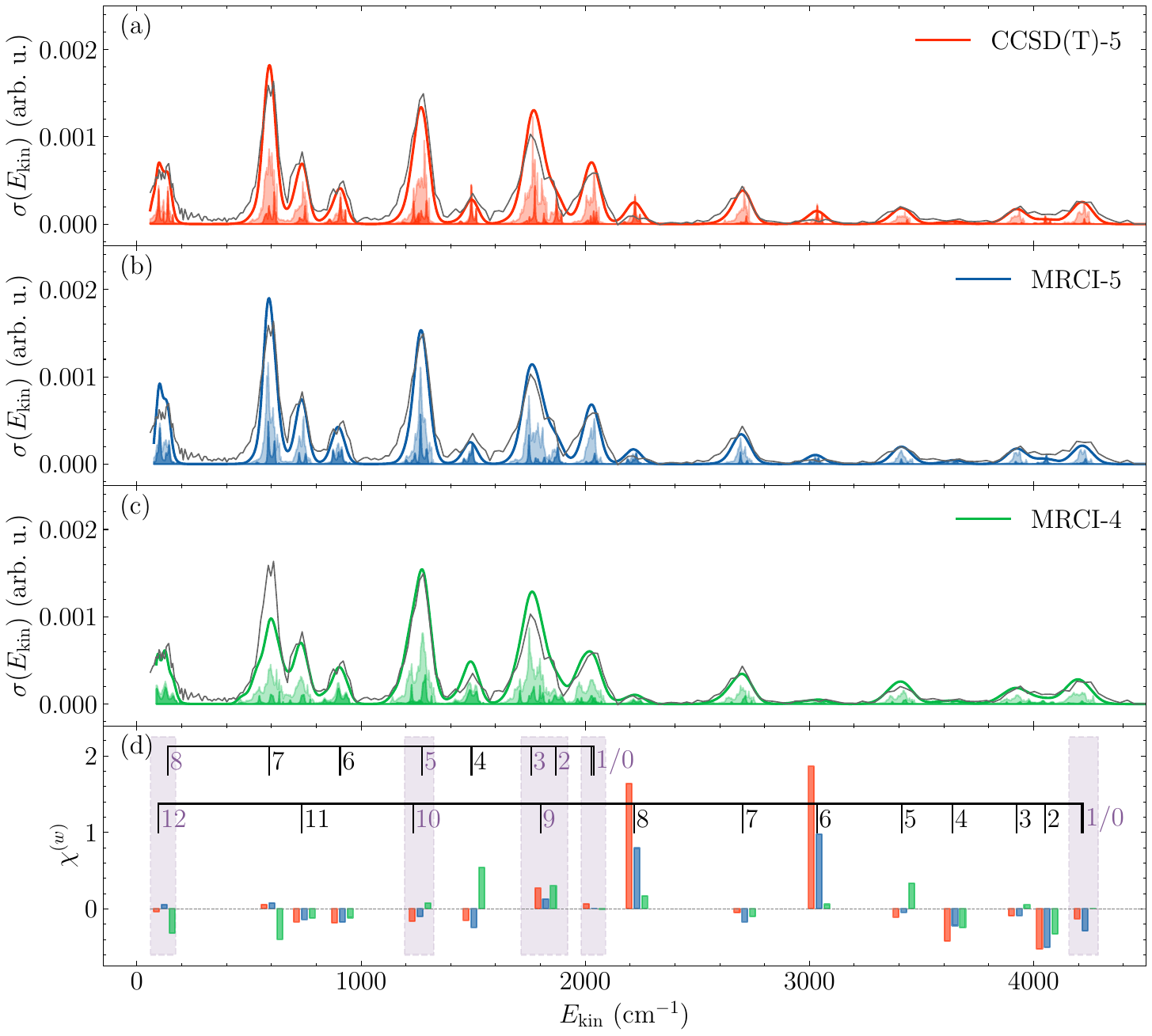}
    \caption{Same as \cref{fig:kinhistov2}, but with the fitted relative scaling
    of \cref{seq:Delta_RMS}.
    }
    \label{sfig:opt_scale_kinhistov2}
\end{figure*}

Experimental noise leads to a certain portion of the
cross section appearing where it should not (i.e., in between peaks,
as can be seen for kinetic energies in the range of $200\,\mathrm{cm}^{-1}$
and $500\,\mathrm{cm}^{-1}$, for instance).
As such, the scaling technique of \cref{eq:cross_section_normalisation} leads
to a slight overestimation of the theoretical peak heights
relative to the experimental ones (because a higher proportion of the
overall area of the cross section curve is concentrated on the peaks).
By simultaneously optimising
for both the scale and the shift in energy according to
\begin{SAlign}
    \label{seq:Delta_RMS}
    \Delta_{\mathrm{opt}} 
    =& \min_{b,E_{\mathrm{shift}}} \Delta(b,E_{\mathrm{shift}}) \\
    =& \min_{b,E_{\mathrm{shift}}} \Big(\int_{\mathbb{R}_+}
        dE_{\mathrm{kin}} \nonumber \\
    &\times  \left|\sigma_{\mathrm{exp}}(E_{\mathrm{kin}}) - b\tilde{\sigma}_{\mathrm{theory}}(E_{\mathrm{kin}}+E_{\mathrm{shift}}) \right|^2\Big)^{-1/2}\,\,, \nonumber
\end{SAlign}
we can find the best possible scale for the simulated cross section in the RMS 
deviation sense.
Above, we have $a=1/\int dE_{\mathrm{kin}}\tilde{\sigma}_{\mathrm{exp}}(E_{\mathrm{kin}})$ and the rescaled theoretical cross section 
is determined by the new, optimised parameter $b$.

The cross sections as a function of kinetic energy 
scaled with the optimised scaling parameter are
shown in \cref{sfig:opt_scale_kinhistov1,sfig:opt_scale_kinhistov2}
for $v=1$ and $v=2$ respectively.
At the cost of additional complexity, these reveal an overall better
fit with the experimental cross section curve in terms of the various
peak heights when compared to their `normalised' counterparts.
Conversely, the peaks in the simulated cross sections with
`normalised scaling' shown in
\cref{fig:kinhistov1,fig:kinhistov2} demonstrate a clear trend of  
overshadowing their measured counterparts, mostly notably for
the $j'=3,5$ and $7$ peaks in the case of the \meuwly and \he PES.
\Cref{tab:regularisation_results} shows the difference that the
chosen scaling scheme makes in terms of the figure of merit $\mathcal{F}$,
(energy-shifted) RMS deviation $\min_{E_{\mathrm{shift}}} \Delta(E_{\mathrm{shift}})$ and optimal energy shift $E_{\mathrm{shift}}^{(\mathrm{opt})}$.
Significantly, for $v=2$, it is seen that in the optimised
scaling scheme, the \meuwly PES outperforms the \ma PES
when it comes to $\min_{E_{\mathrm{shift}}} \Delta(E_{\mathrm{shift}})$.
In contrast, when using the scaling of \cref{eq:cross_section_normalisation},
the energy shifted RMS deviation $\min_{E_{\mathrm{shift}}} \Delta(E_{\mathrm{shift}})$ 
for $v=2$ was minimal for the \ma PES.
In this case the most notable difference is in the overestimation of the height
of the peak comprised of the $v'=1$ and $j'=2,3$ as well as the $v'=0$ and $j'=9$ 
final states, as well as the peak due to $v'=1$ and $j'=7$ for the \meuwly PES.
It is seen that optimised rescaling causes the energy shifted RMS deviation to drop by 
$1.8\cdot10^{-3}$ in the case of \meuwly, whilst the improvement for the \ma PES
is more moderate at $0.3\cdot10^{-3}$.
The figure of merit $\mathcal{F}$ is affected in a similar way, with a larger
margin of improvement observed for \meuwly, causing it to more clearly outperform
the other PES when utilising optimised scaling.

\section{Regularisation}

We have accounted for the presence of noise in the experimental
cross section by investigating whether a regularisation technique,
similar to that employed in \cite{horn2023improving},
influences the results.
Regularisation was implemented by subtracting $\delta_{\mathrm{reg}}$
from the difference in cross sections,
whilst ensuring the kinetic energy-dependent difference never drops below 
zero
\begin{SAlign}
    \epsilon(E_{\mathrm{kin}}) = \max{(|\sigma_{\mathrm{exp}}(E_{\mathrm{kin}})-\sigma_{\mathrm{shift}}(E_{\mathrm{kin}})|-\delta_{\mathrm{reg}},0)}\,\,,
\end{SAlign}
where $\sigma_{\mathrm{shift}}(E_{\mathrm{kin}})$ is the shifted and rescaled simulated cross section under either scaling method.
The non-regularised difference between cross sections in
the expressions for $\Delta(E_{\mathrm{shift}})$ can then be
replaced with the regularised expression $\Delta_{\mathrm{reg}}(E_{\mathrm{shift}})$.
Whilst it was observed that regularisation does indeed affect
the RMS deviation and figure of merit (whose regularised counterpart is 
denoted by $\mathcal{F}_{\mathrm{reg}}$)
as summarised in \cref{tab:regularisation_results}, 
the cross sections due to all PES are affected in almost the same way, i.e., regularisation
does not qualitatively impact the PES comparison results.

\begin{table*}
  \resizebox{\textwidth}{!}{
\begin{NiceTabular}{|c | c | c | c c c  c c|}
\Hline
\Block{1-1}{$v$} & scaling & PES & $\mathcal{F}$ & $\mathcal{F}_{\mathrm{reg}}$ & $\min_{E_{\mathrm{shift}}}\Delta(E_{\mathrm{shift}})$ & $\min_{E_{\mathrm{shift}}}\Delta_{\mathrm{reg}}(E_{\mathrm{shift}})$ & $E_{\mathrm{shift}}^{(\mathrm{opt})}$ \\
\Hline
\Block{6-1}{1}
	&  & CCSD(T)-5 & \colorbox{red!20!white}{$9.9\cdot10^{-3}$} & \colorbox{red!20!white}{$8.0\cdot10^{-3}$} & \colorbox{white}{$9.9\cdot10^{-3}$} & \colorbox{white}{$8.0\cdot10^{-3}$} & \colorbox{red!20!white}{$-0.74\mathrm{cm}^{-1}$}\\
	& Eq. (1) & MRCI-5 & \colorbox{white}{$1.2\cdot10^{-2}$} & \colorbox{white}{$1.0\cdot10^{-2}$} & \colorbox{white}{$1.0\cdot10^{-2}$} & \colorbox{white}{$8.1\cdot10^{-3}$} & \colorbox{white}{$21.0\mathrm{cm}^{-1}$}\\
	&  & MRCI-4 & \colorbox{white}{$1.1\cdot10^{-2}$} & \colorbox{white}{$9.4\cdot10^{-3}$} & \colorbox{green!20!white}{$7.9\cdot10^{-3}$} & \colorbox{green!20!white}{$6.0\cdot10^{-3}$} & \colorbox{white}{$30.0\mathrm{cm}^{-1}$}\\
	\Hline
	&  & CCSD(T)-5 & \colorbox{red!20!white}{$7.4\cdot10^{-3}$} & \colorbox{red!20!white}{$5.1\cdot10^{-3}$} & \colorbox{white}{$7.4\cdot10^{-3}$} & \colorbox{white}{$5.1\cdot10^{-3}$} & \colorbox{red!20!white}{$-0.86\mathrm{cm}^{-1}$}\\
	& Eq. (S1) & MRCI-5 & \colorbox{white}{$1.0\cdot10^{-2}$} & \colorbox{white}{$8.0\cdot10^{-3}$} & \colorbox{white}{$8.1\cdot10^{-3}$} & \colorbox{white}{$5.8\cdot10^{-3}$} & \colorbox{white}{$21.0\mathrm{cm}^{-1}$}\\
	&  & MRCI-4 & \colorbox{white}{$9.9\cdot10^{-3}$} & \colorbox{white}{$8.0\cdot10^{-3}$} & \colorbox{green!20!white}{$6.6\cdot10^{-3}$} & \colorbox{green!20!white}{$4.5\cdot10^{-3}$} & \colorbox{white}{$29.0\mathrm{cm}^{-1}$}\\
	\Hline
\Block{6-1}{2}
	&  & CCSD(T)-5 & \colorbox{red!20!white}{$8.5\cdot10^{-3}$} & \colorbox{red!20!white}{$7.0\cdot10^{-3}$} & \colorbox{white}{$8.5\cdot10^{-3}$} & \colorbox{white}{$7.0\cdot10^{-3}$} & \colorbox{red!20!white}{$-0.54\mathrm{cm}^{-1}$}\\
	& Eq. (1) & MRCI-5 & \colorbox{white}{$1.1\cdot10^{-2}$} & \colorbox{white}{$9.1\cdot10^{-3}$} & \colorbox{white}{$9.7\cdot10^{-3}$} & \colorbox{white}{$8.3\cdot10^{-3}$} & \colorbox{white}{$16.0\mathrm{cm}^{-1}$}\\
	&  & MRCI-4 & \colorbox{white}{$9.0\cdot10^{-3}$} & \colorbox{white}{$7.6\cdot10^{-3}$} & \colorbox{green!20!white}{$7.3\cdot10^{-3}$} & \colorbox{green!20!white}{$6.1\cdot10^{-3}$} & \colorbox{white}{$27.0\mathrm{cm}^{-1}$}\\
	\Hline
	&  & CCSD(T)-5 & \colorbox{red!20!white}{$6.7\cdot10^{-3}$} & \colorbox{red!20!white}{$5.0\cdot10^{-3}$} & \colorbox{red!20!white}{$6.7\cdot10^{-3}$} & \colorbox{red!20!white}{$5.0\cdot10^{-3}$} & \colorbox{red!20!white}{$-0.68\mathrm{cm}^{-1}$}\\
	& Eq. (S1) & MRCI-5 & \colorbox{white}{$7.8\cdot10^{-3}$} & \colorbox{white}{$6.1\cdot10^{-3}$} & \colorbox{white}{$6.9\cdot10^{-3}$} & \colorbox{white}{$5.1\cdot10^{-3}$} & \colorbox{white}{$16.0\mathrm{cm}^{-1}$}\\
	&  & MRCI-4 & \colorbox{white}{$8.6\cdot10^{-3}$} & \colorbox{white}{$7.2\cdot10^{-3}$} & \colorbox{white}{$7.0\cdot10^{-3}$} & \colorbox{white}{$5.7\cdot10^{-3}$} & \colorbox{white}{$27.0\mathrm{cm}^{-1}$}\\
	\Hline
\end{NiceTabular}}
\caption{Expanded version of \cref{tab:results}, including regularisation and the alternative cross section scaling scheme introduced in \cref{seq:Delta_RMS}. For both ``initial'' vibrational states, the PES that  performs best for a given quantifier is highlighted, using the color code of \cref{fig:kinhistov1,fig:kinhistov2}.}
\label{tab:regularisation_results}
\end{table*}


\begin{figure*}[tbp]
\includegraphics[width=0.98\textwidth]{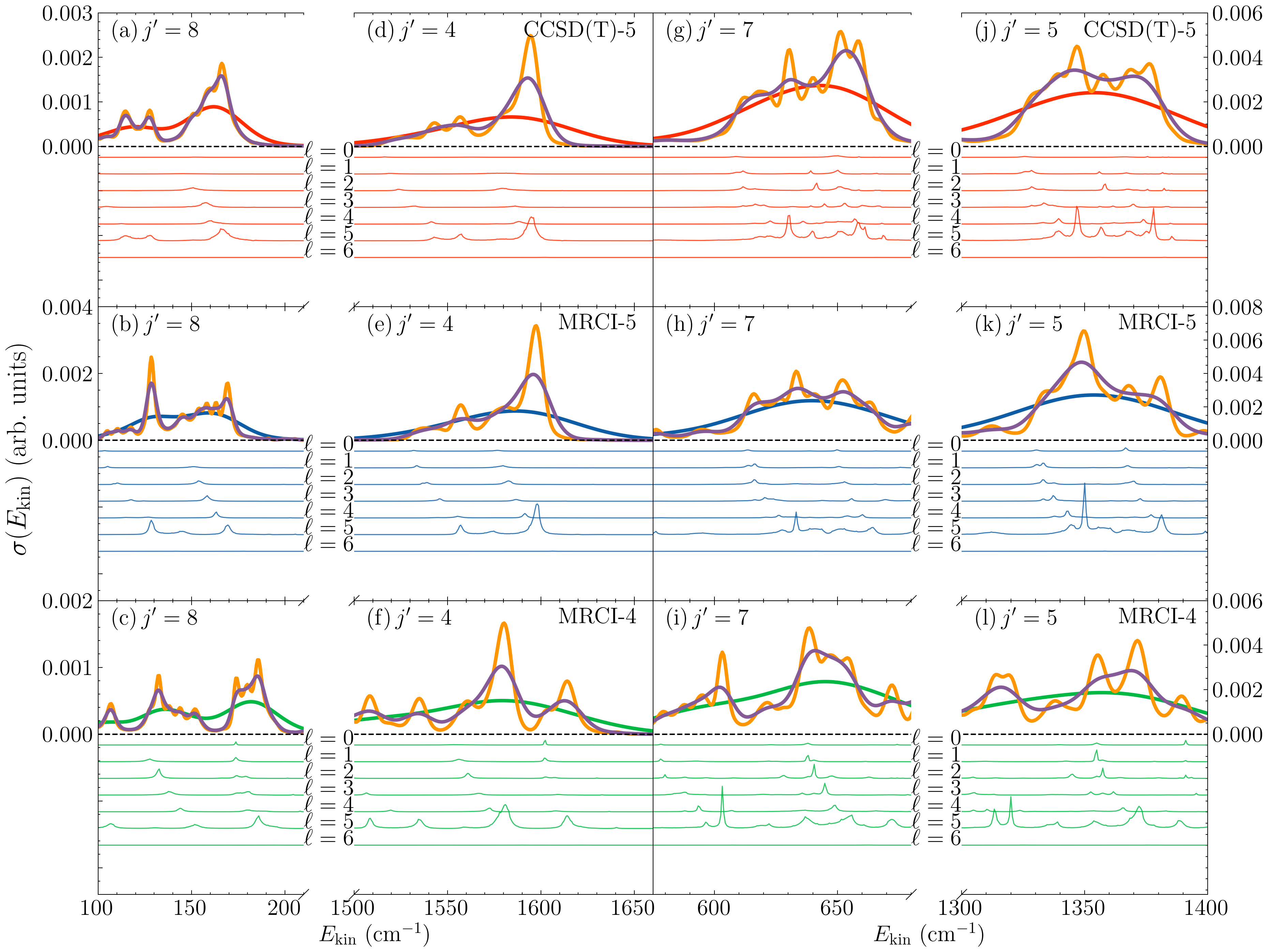}
  \caption{Role of convolution width in the calculated cross sections for $v=1$, 
    using a convolution of fixed width for all energies
    (thick solid  
    lines): Peaks appearing broad in \cref{fig:kinhistov1} 
    at the original convolution width
    start to show substructure when improving
    the (fixed-width) energy resolution by a factor of four
    (purple), resp. 10 (gold). 
    Also shown are the individual contributions of the
    various initial channels (thin solid lines).
    }
    \label{fig:resolution_scaling}
\end{figure*}

\section{Details on convolution resolution scaling}

The decomposition of the
unconvoluted cross section according to the total angular momentum $J$
and initial partial waves $\ell$ (multiplied with -1 and offset for
better visibility) is also shown in \cref{fig:resolution_scaling}:
Contributions with larger $J$ and $\ell$ tend to occur at higher
kinetic energies, most noticeably for para-H$_2^+$.
As such, features present in the gold curves, corresponding to ten times the resolution, can be attributed to a particular $J$ and $\ell$ contribution in many cases.

There are also
initial channels whose contributions cannot be resolved, for example
at $650\wavenum$ for the \meuwly PES in
\cref{fig:resolution_scaling}(g), which corresponds to a total energy
of $-20\wavenum$ in
\cref{fig:cross_section_collision_energy}(d).

\end{document}